\documentclass[superscriptaddress,floatfix,preprintnumbers, nofootinbib,hyperref]{revtex4-2} 
\pdfoutput=1
\usepackage[title]{appendix}
\usepackage[colorlinks=true,breaklinks=true]{hyperref}
\usepackage[normalem]{ulem}
\usepackage[utf8]{inputenc}
\hypersetup{allcolors=[rgb]{0.0 0.0 0.6},linkcolor=[rgb]{0.75 0.05 0.05}}
\usepackage{amsmath,amssymb}
\usepackage{epsfig}  
\usepackage{graphicx}   
\usepackage{slashed}       
\usepackage{url}
\usepackage[dvipsnames]{xcolor}
\usepackage{color}
\usepackage{multirow}
\usepackage{comment}
\usepackage{amssymb}
\usepackage[version=3]{mhchem}
\usepackage{orcidlink}
\usepackage{longtable}

\DeclareUnicodeCharacter{2212}{-}

\allowdisplaybreaks

\bibpunct{[}{]}{,}{n}{}{,}
\definecolor{darkspringgreen}{rgb}{0.09, 0.45, 0.27}

\begin{document}

\vspace*{-1.5 truecm}
\begin{figure}[htbp]
\centering
    \includegraphics[width=4cm]{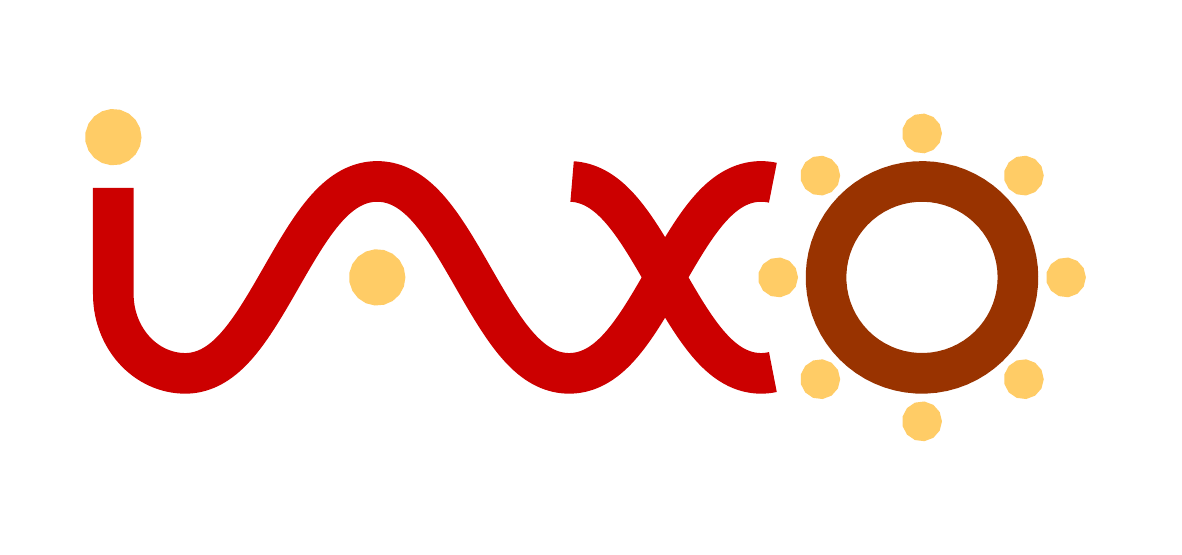}
\end{figure}


\title{The International Axion Observatory (IAXO): case, status and plans.\\ Input to the European Strategy for Particle Physics} 

\author{A. Arcusa}
\affiliation{ITA, Instituto Tecnológico de Aragón, C. María de Luna, 7, 50018 Zaragoza Spain}

\author{S. Ahyoune}
\affiliation{Departament de F\'isica Qu\`antica i Astrof\'isica, Institut de Ci\`encies del Cosmos, Universitat de Barcelona, Barcelona, Spain}

\author{K. Altenm\"uller}
\affiliation{Centro de Astropart\'iculas y F\'isica de Altas Energ\'ias (CAPA), Universidad de Zaragoza, 50009 Zaragoza, Spain}

\author{I. Antol\'in}
\affiliation{Centro de Astropart\'iculas y F\'isica de Altas Energ\'ias (CAPA), Universidad de Zaragoza, 50009 Zaragoza, Spain}
\affiliation{Institute for Experimental Physics, University of Hamburg, Hamburg, 22761, Germany}

\author{S. Basso}
\affiliation{INAF, Osservatorio Astronomico di Brera, via Bianchi 46, 23807 Merate (LC), Italy}

\author{P. Brun}
\affiliation{IRFU, CEA, Universit\'e Paris-Saclay, F-91191 Gif-sur-Yvette, France}

\author{V. Burwitz}
\affiliation{Max-Planck-Institut für extraterrestrische Physik, Giessenbachstr., 85748 Garching, Germany}

\author{F. R. Cand\'on}
\affiliation{Centro de Astropart\'iculas y F\'isica de Altas Energ\'ias (CAPA), Universidad de Zaragoza, 50009 Zaragoza, Spain}

\author{J. F. Castel}
\affiliation{Centro de Astropart\'iculas y F\'isica de Altas Energ\'ias (CAPA), Universidad de Zaragoza, 50009 Zaragoza, Spain}

\author{S. Cebri\'an}
\affiliation{Centro de Astropart\'iculas y F\'isica de Altas Energ\'ias (CAPA), Universidad de Zaragoza, 50009 Zaragoza, Spain}

% ...


\begin{abstract}
The International Axion Observatory (IAXO) is a next-generation axion helioscope designed to search for solar axions with unprecedented sensitivity. IAXO holds a unique position in the global landscape of axion searches, as it will probe a region of the axion parameter space inaccessible to any other experiment. In particular, it will explore QCD axion models in the mass range from meV to eV, covering scenarios motivated by astrophysical observations and potentially extending to axion dark matter models. 
Several studies in recent years have demonstrated that IAXO has the potential to probe a wide range of new physics beyond solar axions, including dark photons, chameleons, gravitational waves, and axions from nearby supernovae.
IAXO will build upon the two-decade experience gained with CAST, the detailed studies for BabyIAXO, which is
currently under construction, as well as new technologies.
If, in contrast to expectations, solar axion searches with IAXO ``only'' result in limits on new physics in presently uncharted parameter territory, these exclusions would be very robust and provide significant constraints on models, as they would not depend on untestable cosmological assumptions.
\end{abstract}
\smallskip

\maketitle

\newpage 

\section{Introduction}
\label{sec:Introduction}




Although axions have been experimentally searched for since their introduction in the 1970's, it is only in recent years that this field is experiencing a strong increase of interest in the world-wide community. 
Exciting new detection concepts are being proposed and many of them are being actively explored in small experimental setups. 
At the same time, the field is experiencing consolation as detection strategies are proved and the community looks towards the next generation of searches to access new parts of the parameter space. This naturally drives the need for larger facilities, of a scale generally not found in the axion community. One of these efforts is the International Axion Observatory (IAXO). The international context of the experiment is provided in other documents to the European Strategy for Particle Physics (ESPP) update process, in particular the one produced by the COSMIC WISPers COST action~\cite{Aybas:2025pbq},
as well as the Physics Beyond Colliders Study Group at CERN~\cite{espp_pbc}. 
\section{Physics case and context}
A very motivated category of extensions of the Standard Model (SM) predicts particles that could lie hidden at the \textit{low energy frontier} (i.e., very light particles), of which the axion is the prototype. Axions naturally appear in models that include the Peccei-Quinn (PQ) mechanism~\cite{Peccei:1977hh,Peccei:1977ur,Weinberg:1977ma,Wilczek:1977pj}, that offers the most promising solution to one of the most serious problems of the Standard Model to date: the \textit{strong-charge-parity (-CP) problem}, or why the strong interactions do not seem to violate the CP symmetry (while according to quantum-chromodynamics they are expected to). Very light and very lightly coupled axion-like particles (ALPs) generically emerge in extensions of the SM with spontaneous symmetry breaking of new symmetries at higher energies~\cite{Jaeckel2010}. Notably, string theory is known to predict many such ALP fields, including the axion itself~\cite{Svrcek:2006yi}.

Being very light particles, axions and ALPs could affect stellar evolution in a way similar to neutrinos, and play important roles in diverse cosmological scenarios. Most relevantly, these particles constitute excellent cold Dark Matter (DM) candidates. 
Despite their low mass, non-relativistic axions can be produced in the early Universe by non-thermal mechanisms like vacuum-realignment and the decay of topological defects of the axion field~\cite{Bae:2008ue,Visinelli:2011wa}. 
The computation of the relic axion density $\Omega_a$ for a given axion model (and thus the ``prediction" of the needed axion mass to obtain the observed DM density) is rather uncertain, from both a cosmological and computational perspective. 
As a result, we remain with a fairly large range of masses to search for the DM axion. Typically a mass range of about $10^{-6}-10^{-3}$~eV is quoted as the cosmologically preferred one, although masses outside this range are also possible  (i.e. they may provide a value for $\Omega_a$ similar to the measured total dark matter density). 
In particular, values above $10^{-3}$~eV and up to 0.1~eV are preferred in models with long-lived topological defects~\cite{Ringwald:2015dsf}. 
These models are appealing because they move the DM window into the astrophysically motivated region. 
Indeed, a few-meV axion is a crucial section of the ALP parameter space, in which motivations from theory (strong CP), cosmology (DM) and astrophysics overlap. 
IAXO will be uniquely sensitive to this region. In addition, if the axions make up only a subdominant part of DM, the right axion mass moves always to higher values with an approximate factor $m_a \sim \Omega_a^{-1}$.

More generic ALPs can also be the DM in a wider range of parameters~\cite{Arias:2012az}. Independently of the relevance to the origin of DM, ALPs have been invoked in different cosmological scenarios. They could constitute extra relativistic degrees of freedom in the early Universe~\cite{Turner:1986tb} (i.e. Dark Radiation, whose presence could be --slightly-- favoured by recent CMB observations~\cite{Bernal:2016gxb}). These dark radiation axions/ALPs can feature a coupling to photons which is likely to be in the IAXO range and can thus be observed independently by IAXO for most values of the relevant parameters~\cite{Ferreira:2018vjj,DEramo:2018vss}. In addition, recent models (dubbed ``ALP miracle'') in which the ALP is identified with the inflaton while, at the same time, composing the DM, have produced rather well-defined predictions that could be testable by IAXO~\cite{Daido:2017tbr,Daido:2017wwb}. Finally, particular realizations of ALP fields (e.g. chamaleon or galileon theories) may constitute viable particle interpretations of Dark Energy.

\begin{figure}[t]
    \centering
    \includegraphics[width=0.8\linewidth]{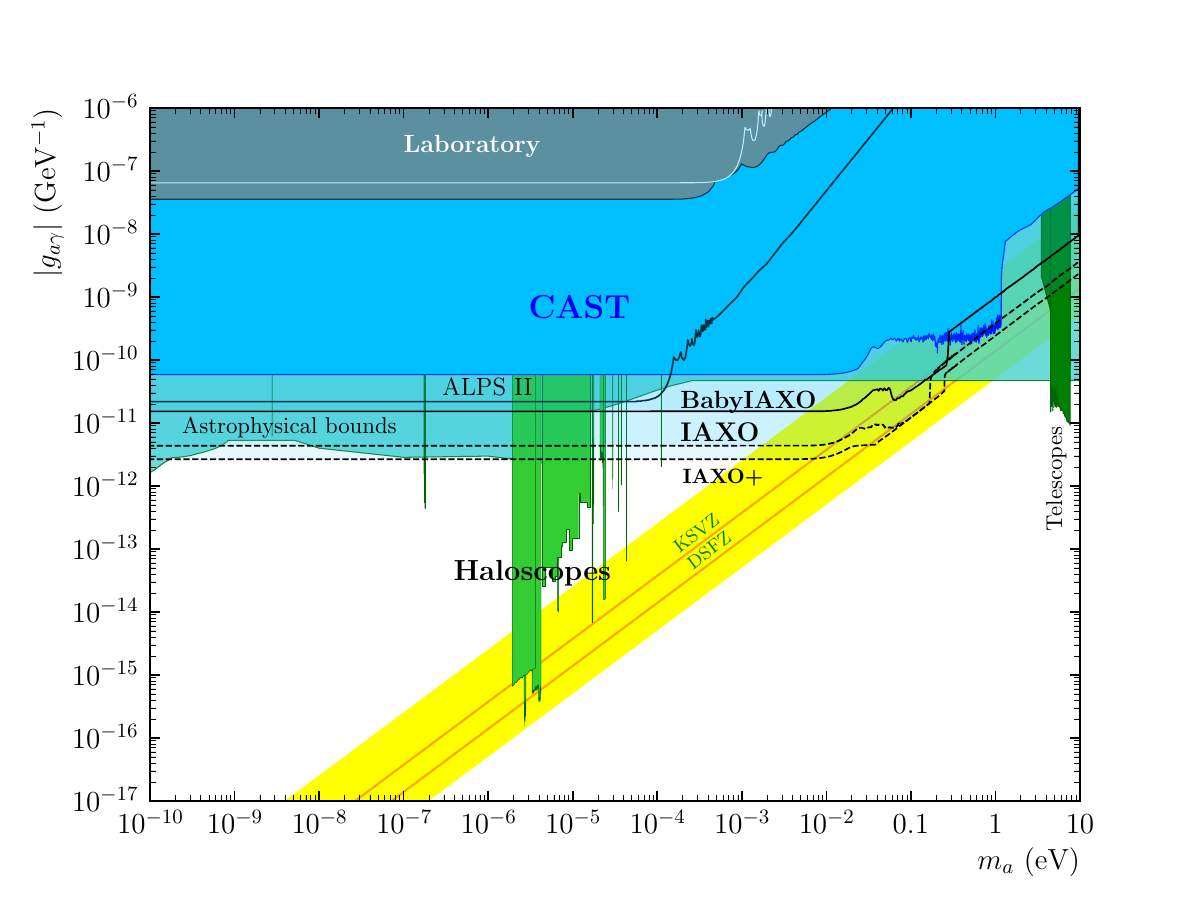}
    \caption{
    ALP parameter space in the region of interest for IAXO, assuming only the coupling to photons. 
    For comparison, we show also other phenomenological and experimental limits, in addition to the predicted sensitivities of IAXO and IAXO+.
    The current bound from CAST is shown in sky blue,
    laboratory experiment bounds in gray, astrophysical bounds in turquoise, and astrophysical searches for axions as dark matter in green.
    The yellow shaded band represents the preferred region for the QCD axion~\cite{DiLuzio:2016sbl,DiLuzio:2021ysg,Plakkot:2021xyx}.    }
    \label{fig:axion-parameter-space}
\end{figure}

Astrophysics has been exhaustively used to constrain the properties of axions and ALPs~\cite{Carenza:2024ehj}. Despite more than 35 years of efforts, astrophysical constraints, however, still leave a relatively large window for the existence of axions (in brief, the axion-photon coupling $g_{a\gamma}$ must be lower than $\sim 10^{-10}$~GeV$^{-1}$ and the axion mass $m_a$ lower than $\sim 1$~eV). Intriguingly, some astrophysical observations actually seem to hint at the presence of an axion or ALP. On one hand, the Universe seems to appear too transparent to very high-energy photons~\cite{DeAngelis:2011id,Horns:2012fx,Aharonian:2005gh,Aliu:2008ay,Teshima:2007zw}, something that has prompted several authors~\cite{Horns:2012fx,Meyer:2013pny,Csaki:2003ef,DeAngelis:2008sk,Roncadelli:2008zz,Simet:2007sa,SanchezConde:2009wu,Dominguez:2011xy,DeAngelis:2011id,Tavecchio:2012um,Rubtsov:2014uga,Kohri:2017ljt} to suggest explanations involving photon-ALP oscillations triggered by cosmic magnetic fields. For this solution to work the required ALP mass must be $m_a \lesssim 10$~neV and its coupling to photons $g_{a\gamma} \sim 10^{-11}$~GeV$^{-1}$. Of course, more standard explanations or systematic effects cannot be ruled out at the moment. In any case, the ALP solution to this anomaly will be fully tested by IAXO. On the other hand, an excessive cooling rate is measured in many stars at different evolutionary stages: red giants, supergiants, helium core burning stars, white dwarfs, and neutron stars. Collectively observations are in $>3\sigma$ tension with stellar models, suggesting a new energy loss mechanism could be at work. Interestingly, a QCD axion of few-meV would provide a perfect fit~\cite{Giannotti:2017hny,Giannotti:2015kwo}. Most of the parameter space invoked by these hints will be at reach of IAXO. In particular, a few meV axion is currently only realistically testable by IAXO, as shown in Figure~\ref{fig:axion-parameter-space}.

The main axion detection strategies make use of strong laboratory magnetic fields to trigger their conversion into detectable photons~\cite{Irastorza:2018dyq}. 
In principle axions can be searched for purely at the laboratory,
e.g. by light-shining-through-wall configurations. However, at present the tiny conversion factors involved preclude sensitivities to vanilla QCD axion models. 
Natural sources of axions are therefore needed. If axions compose most of our galactic DM halo, they could produce detectable signals in appropriately designed detectors. The \textit{axion haloscope}~\cite{Sikivie:1983ip} is the most famous technique to search for DM axions. 
It has been proven to be able to reach sensitivity to relevant QCD models in the mass range of 2--10~$\mu$eV, and there are ongoing searches with discovery potential here, e.g. by ADMX.
Outside this range, the technique still requires significant developments. 
A number of initiatives to extend the axion haloscope technique, or to test altogether novel detection concepts, have recently been proposed to target DM axions in different mass points. As a result, a diverse and vibrant landscape of experimental efforts is emerging, although most of them are constrained to relatively small R\&D setups. 
However, it has become increasingly clear in recent years that the prospects for accurately determining the local dark matter density on solar-system scales are rather limited in the coming decades~\cite{Kim:2024xcr}. As a consequence, though axion dark matter searches continue to offer remarkable discovery potential, null results remain difficult to interpret.

\textit{Axion helioscopes} looking for axions emitted by the Sun represent a particular category of experiments. 
It is the only approach that combines relative immunity to model assumptions (solar axion emission is a generic prediction of most axion models) plus a competitive sensitivity to axion parameters largely complementary to those accessible with other detection techniques. The current state-of-the-art is represented by the CERN Axion Solar Telescope (CAST), which however stopped data taking some years ago. 
Its latest results~\cite{CAST:2024eil}, shown in Figure~\ref{fig:axion-parameter-space}, are comparable with the best astrophysical limits on the axion-photon coupling ($g_{a\gamma}$) for a wide axion mass ($m_a$) range. IAXO aims at going substantially beyond CAST sensitivity and therefore deeply entering into unexplored regions. As shown in Figure~\ref{fig:axion-parameter-space}, the region of ALP parameter space searchable by IAXO is largely complementary to the one accessible to haloscopes, and features a different set of assumptions. In particular, helioscopes do not rely on the axion being the DM. 

In addition, IAXO might also be sensitive to solar axions produced by mechanisms mediated by the axion-electron coupling $g_{ae}$ or alternatively produce the most stringent limits on this coupling. In the case of an axion discovery, IAXO might provide insight on both the axion electron and the axion photon couplings, providing critical information about the nature of the underlying axion model~\cite{Jaeckel:2018mbn}. Similarly, the determination of the axion’s coupling to nucleons could be achieved by observing specific emission lines from nuclear de-excitation, such as those from $^{57}$Fe~\cite{DiLuzio:2021qct}.
A positive detection in IAXO would also allow: 1) the determination of its intrinsic parity via photon polarisation measurement, 2) in some circumstances, the determination of the axion mass, via the spectral distortions due to the loss of coherence~\cite{Dafni:2018tvj}, 3) to open a new window to solar observation, e.g. examining the internal solar magnetic field~\cite{OHare:2020wum}, the  internal solar temperature profile~\cite{Hoof:2023jol}, or its metallicity ~\cite{Jaeckel:2019xpa}, potentially solving the solar abundance problem.

It is worth pointing out that the large magnetic volume of the IAXO  (and BabyIAXO) magnet could also be used for additional types of searches, effectively turning IAXO into a versatile infrastructure for axion/ALP searches. 
Most prominently, there is the possibility of  using IAXO as an \textit{axion haloscope}, an approach currently being explored by the RADES collaboration~\cite{Ahyoune:2023gfw}, pointing to very competitive prospects for BabyIAXO to search for dark matter axions, especially at the 1-2 $\mu$eV scale. 
With this goal in mind, the RADES collaboration already released a number of physics results~\cite{Ahyoune:2024klt,CAST:2020rlf} to build up the expertise in this detection technique.
The ongoing effort is further supported by the \textit{DarkQuantum} ERC-SyG~\cite{DarkQuantum} and QRADES~\cite{Quantera} grants, which enable the pursuit of this goal using quantum-limited detection methods.

Another emerging topic is the possibility of searching for high-frequency gravitational waves (HFGW)~\cite{Aggarwal:2025noe} with axion-inspired experiments. 
The absence of known astrophysical sources above $\sim 10$kHz turns  this frequency range into a particularly clean window to search for signals of physics beyond the Standard Model.
Further studies are needed to refine the sensitivity to HFGWs, but it is already clear that significant synergies can be exploited. Initial investigations of the potential of (Baby)IAXO in this context have already been presented~\cite{Ringwald:2020ist,Valero:2024ncz}.  


Another interesting discovery opportunity for IAXO would be the detection of Supernova (SN) axions~\cite{Carenza:2025uib}. A sufficiently close SN event ($d \lesssim 300$ pc) could produce a flux high enough to be detected by IAXO (and, to a lesser extent, by BabyIAXO), provided they are equipped with an appropriate MeV detector. 
This measurement could yield critical information about the equation of state of matter under extreme conditions, as well as extend the exploration of axion parameter space in regions that are inaccessible to solar observations.

To summarize, the physics case of IAXO can be condensed in the following few statements:

\begin{itemize}
\item IAXO follows the only proposed technique able to probe a large fraction of QCD axion models in the meV to eV mass band (e.g. KSVZ axions down to a few meV will be probed). \textbf{No other proposed experiment provides sensitivity in this region}. This is a crucial area of the axion parameter space, in which the astrophysical, cosmological (DM) and theoretical (strong CP problem) motivations overlap. 

\item IAXO will fully probe the ALP region invoked to solve the transparency anomaly, and will largely probe the axion region invoked to solve observed stellar cooling anomalies. 


\item The above sensitivity goals do not depend on the hypothesis of axions being the DM, i.e. in case of non-detection, IAXO will robustly exclude the corresponding range of parameters for the axion/ALP.


\item In case of a positive detection, IAXO will offer a rich post-discovery program, including confirmation of intrinsic parity, information on the nature of the underlying axion model (i.e. determination of other couplings), possibly also determination of the axion mass, opening a new window of solar observation. 

\item Finally, IAXO will also constitute a generic infrastructure for axion/ALP physics with potential for additional  search strategies, like the search for dark matter axions (operating as an axion haloscope), high-frequency gravitational waves or Supernova axions. 

\end{itemize}

\section{The IAXO facility}

\begin{figure}[t!] \centering
\includegraphics[trim= {20cm 0 20cm 0}, clip, height=0.35\textwidth]{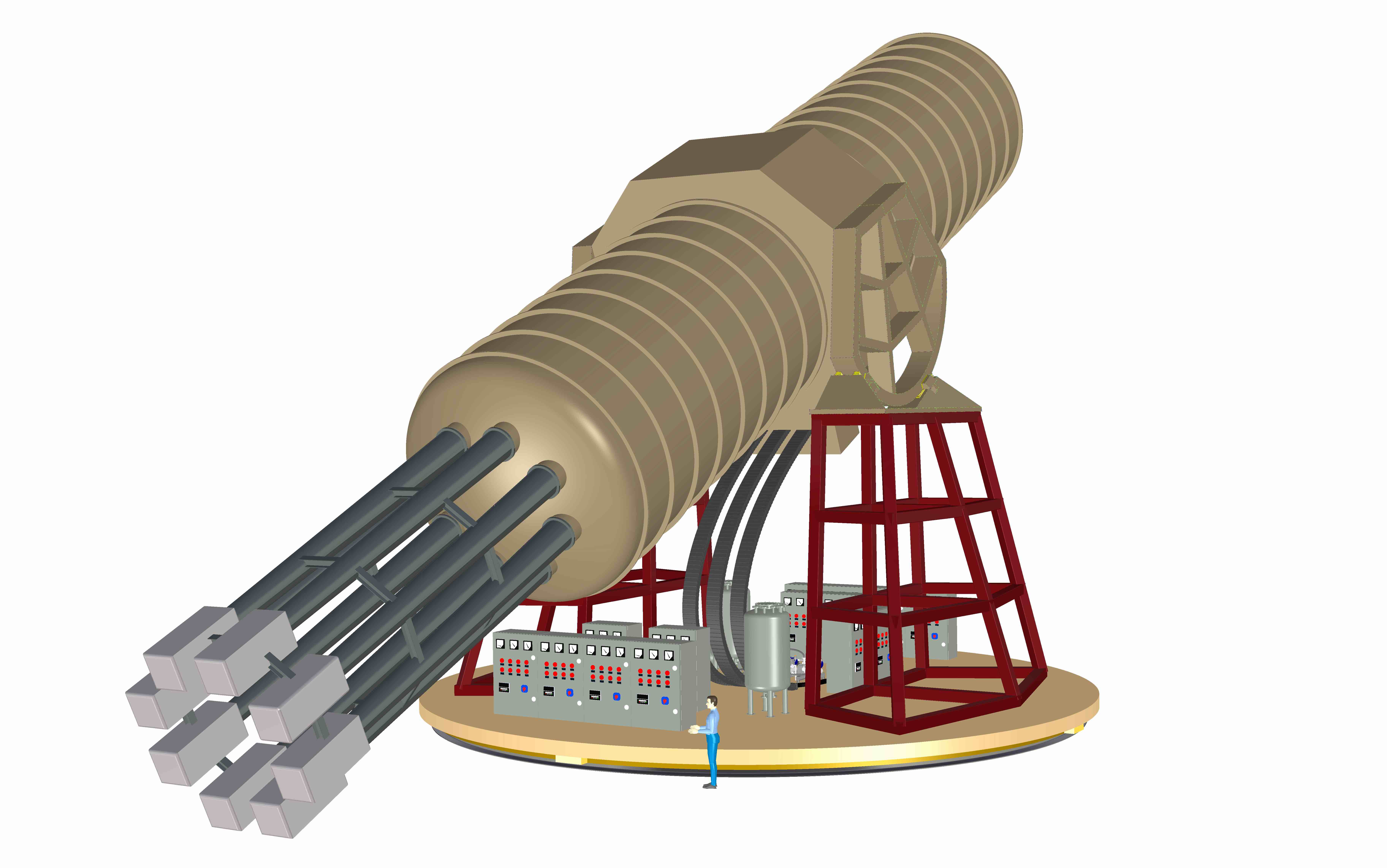}%
\includegraphics[trim= {6cm 4cm 6cm 5cm}, clip, height=0.275\textwidth]{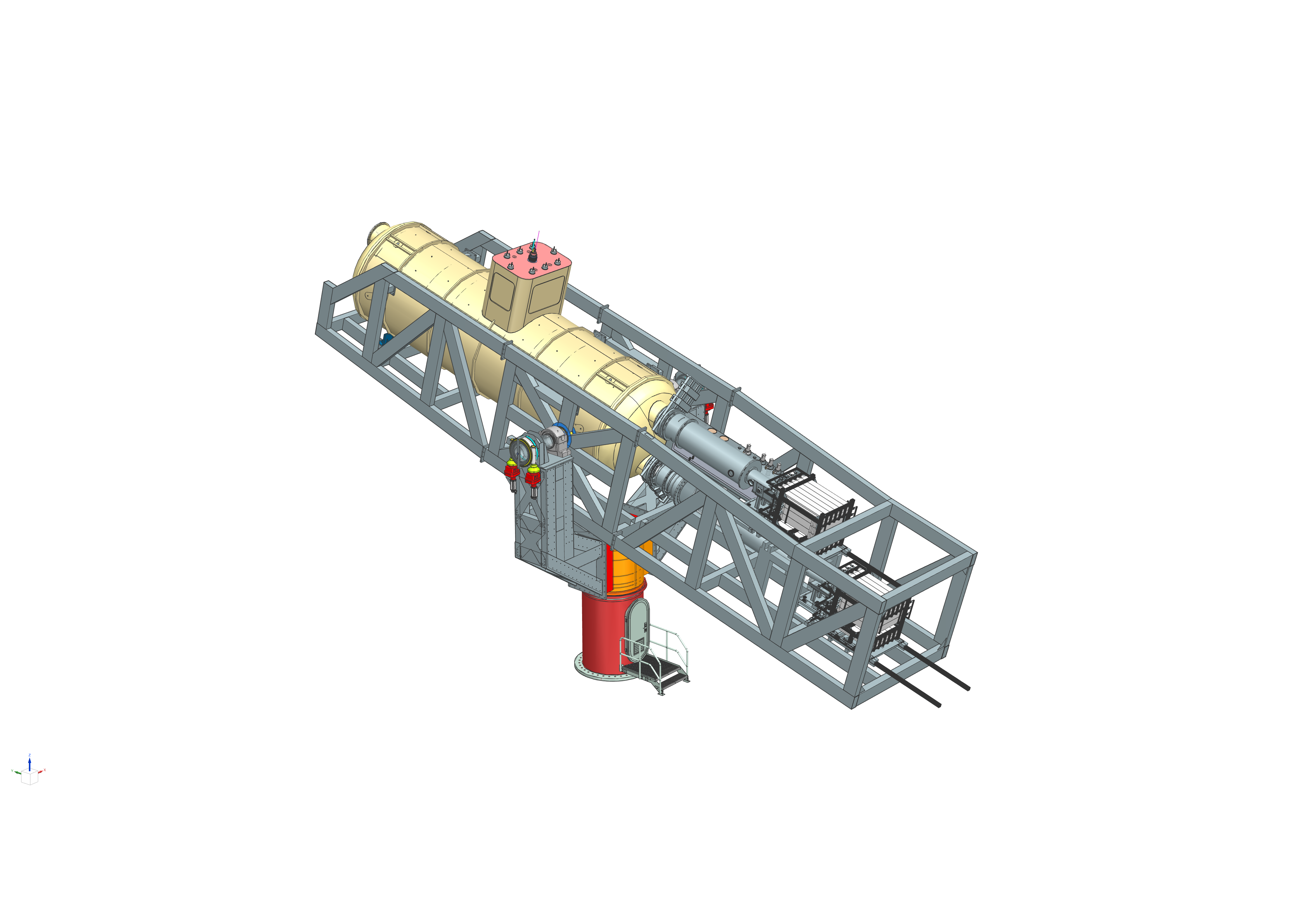}%
\caption{\label{fig:IAXO_cdr} \textit{Left:} Conceptual design of IAXO as defined in~\cite{Armengaud:2014gea}. \textit{Right:} Most recent CAD Design of BabyIAXO.}
\end{figure}

The next-generation axion helioscope IAXO will achieve an improvement of more than 4-5 orders of magnitude in signal-to-noise ratio compared to CAST, currently the most sensitive experiment, translating into sensitivity to the axion-photon coupling $g_{a\gamma}$ down to a few $10^{-12}$~GeV$^{-1}$—roughly a factor of 20 better than the current best limit set by CAST. 
This leap in overall sensitivity is achieved by (1) the realization of a large-scale magnet, (2) extensive use of X-ray focusing optics, (3) low background detectors and (4) enhanced tracking capabilities increasing the total exposure in axion-sensitive conditions. Figure~\ref{fig:IAXO_cdr} shows a view of the experiment. 

The main element of IAXO is thus a new dedicated large superconducting magnet, designed to maximize the helioscope figure of merit. The IAXO magnet will follow a large multi-bore toroidal configuration, to efficiently produce an intense magnetic field over a large volume. The design is inspired by the ATLAS barrel and end-cap toroids, the largest superconducting toroids ever built and presently in operation at CERN. Indeed, the experience of CERN in the design, construction and operation of large superconducting magnets is a key aspect of the project. 

The X-ray focusing relies on the fact that it is possible to fabricate X-ray mirrors with high reflectivity at grazing incident angles. IAXO envisions newly-built optics similar to those used onboard NASA's NuSTAR satellite mission~\cite{nustar2013}, but optimized for the energies of the solar axion spectrum and adapted in size to match the envisioned IAXO dimensions. Other technologies, like the Al-foil-based optics of the XRISM mission are also contemplated. Each of the eight $\sim$60~cm diameter magnet bores will be equipped with such optics. At the focal plane of each of the telescopes, IAXO will have low-background X-ray detectors. Several detection technologies are under consideration, but the most developed types are small gaseous chambers read by pixelised microbulk Micromegas planes~\cite{Andriamonje:2010zz,Aznar:2015iia,Altenmuller:2024uza}. They involve low-background techniques typically developed in underground laboratories, like radiopure detector components, appropriate shielding, and offline discrimination algorithms. Alternative or additional X-ray detection technologies are also considered, like GridPix detectors, Magnetic Metallic Calorimeters, Transition Edge Sensors, or Silicon Drift Detectors. All of them show promising prospects to outperform the baseline Micromegas detectors in aspects like energy threshold or resolution, which are of interest, for example, to search for solar axions via the axion-electron coupling, a process featuring both lower energies than the standard Primakoff ones, and monochromatic peaks in the spectrum.  Additional hardware is considered to extend the physics case along the lines presented above, like RF-cavities and sensors for dark matter axions or MeV detectors for SN axiones.

\hyphenation{Baby-IAXO}

The collaboration is currently building an intermediate stage, BabyIAXO, described in the next section. The experience with BabyIAXO and its physics results will impact the design of the final IAXO. We use the label ``IAXO+'' to represent a possible enhanced FOM for IAXO, somewhat arbitrarily fixed as an improved factor of 10 with respect to the nominal set of design parameters, which would enable probing most of the astrophysically motivated models.
Both scenarios, nominal and enhanced, are represented in the sensitivity figure~\ref{fig:axion-parameter-space}. The plans of the collaboration is to revisit the design of IAXO with the experience gathered in the construction of BabyIAXO, and improve its FOM, potentially making the 
``IAXO+'' scenario our new nominal one.

\section{BabyIAXO and near term plans}
\hyphenation{Baby-IAXO}

An intermediate experimental stage called BabyIAXO is currently under construction by the collaboration. BabyIAXO will test magnet, optics and detectors at a technically representative scale for the full IAXO, and, at the same time, it will be operated and will take data as a fully-fledged helioscope experiment, with sensitivity beyond CAST and potential for discovery. 

BabyIAXO will feature a 10~m long superconducting magnet following a ``common coil'' layout, i.e. two flat racetracks coils that resemble the unit coils of the final IAXO magnet described in the previous section. In between the coils two parallel 10~m long, 70~cm diameter bores are placed. Both bores will be equipped with detection lines (optics and detector) of similar dimensions as the final ones foreseen for IAXO. The overall experiment is shown on the right of Figure~\ref{fig:IAXO_cdr}. 

The baseline X-ray optic for BabyIAXO is an X-ray telescope that is as close as possible in dimensions and performance to the final IAXO optic, based on the same technology, and made of multiple thermoformed substrates. 
In order to fabricate the BabyIAXO optic quickly and cost-efficiently, the production leverages investments of NASA's NuSTAR mission in both technology and construction~\cite{10.1117/12.826724}, as well as cold glass replication techniques~\cite{2016SPIE.9905E..6UC}, taking advantage of a purpose-designed integration machine at INAF. Additionally to one of the IAXO baseline optics, the collaboration is planning on using an existing, $70$-cm diameter XMM Newton~\cite{Jansen:2001bi} flight-spare optic on the second BabyIAXO magnet bore. Two such optics exist and they are property of ESA. 
A loan agreement is with ESA to use one such optics in BabyIAXO is in preparation.

In addition to being a technological demonstrator of all the subsystems of IAXO, BabyIAXO will enjoy a relevant physics potential in itself, and will allow the collaboration to move into ``experiment" mode relatively early, and, in doing so, create intangible resources in the collaboration (working groups on analysis, software, data taking, etc.) that will be needed for IAXO. The experimental parameters expected from the above design allow BabyIAXO to probe new unexplored axion and ALP parameter space, as shown in Figure~\ref{fig:axion-parameter-space}.

\section{From BabyIAXO to IAXO}


IAXO currently brings together around 125 researchers from 24 institutions, primarily based in Europe and the United States, including key partners such as CERN and DESY. The present focus is on the construction of BabyIAXO, with particular emphasis on the development of its superconducting magnet—an area where CERN’s expertise is especially critical. The experiment will be hosted at DESY and could begin solar axion searches as early as 2029.

 
The magnet of IAXO is of a size and field strength comparable to that of large detector magnets typically built in high energy physics. For this, IAXO relies on the unique expertise of CERN in large superconducting magnets. The CERN magnet detector group has already led all the magnet design work so far for the IAXO CDR~\cite{Armengaud:2014gea}. 
While the baseline concept for the IAXO magnet relies on an aluminium stabilized Rutherford cabel, the use of High Temperature Superconducting (HTS) material is becoming an interesting alternative. HTS offers the possibility of operating the system at higher temperature and generating higher magnetic fields. The advantages would be a reduced cost of the cryogenic cooling - helium gas at any temperature in the range from 20 K to about 60 K could be the coolant - and an higher temperature margin during operation. The HTS material would be the REBCO tape, today available on the market in reasonably long lengths (200 m to 600 m) and at an affordable price. The project could benefit from the synergy with other on-going activities, for instance the High Field Magnet HTS activity at CERN where REBCO racetrack coils and cables are being developed and tested \cite{Ballarino2025}
A demonstrator REBCO coil, with representative dimensions, is proposed to be developed and qualified prior to launching the design of the final magnet.  The project would benefit from contributions from other international laboratories.
Realizing the IAXO magnet with HTS might also trigger other larger scale magnet projects based on this modern conductor type.


\section{BabyIAXO and IAXO at DESY}

In DESY's longer term strategy for future particle physics experiments, IAXO is a very prominent option, which would ideally complement DESY's ongoing ALPS II activities and its participation in remote accelerator based experiments. Therefore, DESY has offered to contribute already in the preparatory phase with support from its project office to structure and coordinate the international activities. In addition, options for BabyIAXO and IAXO sites on the DESY campus in Hamburg are presently being investigated in detail. BabyIAXO  could be located in one of the big experimental halls of the HERA facility, which was shut down in 2007. Their infrastructure is well suited to construct and operate the prototype.   
In October 2018, BabyIAXO and IAXO have been presented to the DESY Physics Research Committee (PRC), a panel of external experts advising the DESY directorate on all matters related to the particle physics programs at DESY. The PRC has welcomed the proposal and 
is regularly following and evaluating the project progress since then.
As IAXO will require significant international funding, 
the outcome of the update process of the ESPP will provide important input to define the next steps.

\section{Conclusions}

Axion searches are now recognized as 
one of the most compelling portals to explore new physics beyond the Standard Model. IAXO has a privileged position in the vibrant world-wide landscape of axion experiments. 
The axion helioscope technique is the only realistic strategy to explore certain well motivated regions of the parameter space, in particular QCD axion models with masses between meV and eV. 
IAXO will test a significant portion of the astrophysically motivated regions, including specific axion DM models. 
This potential is largely complementary to other efforts of the community, as stressed in respective community documents to the ESPP~\cite{Aybas:2025pbq}

As has been detailed in this document, IAXO is solidly based on previous experience in the CAST experiment, as well as on a number of well-proven solutions and technologies, most relevantly on superconducting magnet expertise present at CERN. 
The IAXO collaboration
is the largest collaboration in experimental axion research, and encompasses all the needed knowledge and capabilities to carry out the IAXO program. 
The near term goal of the collaboration is to build and operate BabyIAXO, an intermediate stage that will serve as technological pathfinder and will already 
provide significant BSM discovery potential. 
Its construction could start in early 2026, pending decisions on funding proposals.
BabyIAXO will enable the collaboration to quickly move into ``science mode'' and will lay the groundwork for the full IAXO. 

IAXO has the potential to become a flagship project of the international axion community. 
Although still relatively small compared to the large HEP facilities, the final IAXO experiment will be of an unprecedented scale for typical axion experiments. 
The consideration of axion searches by the European Strategy for Particle Physics, and in particular of the IAXO project, is therefore required to face the challenges of its realization in the coming decade.


\bibliographystyle{bibi.bst}
\bibliography{references.bib}

\end{document}